\documentclass[fleqn,10pt]{wlscirep}
\usepackage[utf8]{inputenc}
\usepackage{amsthm}
\usepackage{algorithm2e}
\title{Underestimated cost of targeted attacks on complex networks}

\author[1]{Xiao-Long Ren}
\author[1]{Niels Gleinig}
\author[2]{Dijana Toli{\'c}}
\author[1,*]{Nino Antulov-Fantulin}
\affil[1]{Computational Social Science, ETH Z\"urich, Clausiusstra{\ss}e 50, 8092 Z\"urich, Switzerland}
\affil[2]{Laboratory for Machine Learning and Knowledge Representations, Rudjer Bo{\v s}kovi\'c Institute, Zagreb, Croatia}

\affil[*]{anino@ethz.ch}

\begin{abstract}
The robustness of complex networks under targeted attacks is deeply connected to the resilience of complex systems, i.e., the ability to make appropriate responses to the attacks.
In this article, we investigated the state-of-the-art targeted node attack algorithms and demonstrate that they become very inefficient when the cost of the attack is taken into consideration. In this paper, we made explicit assumption that the cost of removing a node is proportional to the number of adjacent links that are removed, i.e., higher degree nodes have higher cost. 
Finally, for the case when it is possible to attack links, we propose a simple and efficient edge removal strategy named Hierarchical Power Iterative Normalized cut (HPI-Ncut).
The results on real and artificial networks show that the HPI-Ncut algorithm outperforms all the node removal and link removal attack algorithms when the cost of the attack is taken into consideration. In addition, we show that on sparse networks, the complexity of this hierarchical power iteration edge removal algorithm is only $O(n\log^{2+\epsilon}(n))$.
\end{abstract}
\begin{document}

\flushbottom
\maketitle

\thispagestyle{empty}

\section{Introduction}
Resilience of complex networks refers to their ability to react on internal failures or external disturbances (attacks) on nodes or edges.
The reaction is fundamentally connected to the robustness of the network structure \cite{Newman2003} that represents the complex system, which is often characterized by the existence of a giant connected component (GCC).
Robustness of the connected components under random failure of nodes or edges is described with classical percolation theory \cite{Sethna06statisticalmechanics}. In network science, percolation is the simplest process showing a continuous phase transition, scale invariance, fractal structure and universality and it is described with just a single parameter, the probability of removing a node or edge.
Network science studies have demonstrated that scale-free networks \cite{BA, Dorogovtsev2000} are more robust than random networks \cite{ER,Gilbert1959} under random attacks but less robust under targeted attacks \cite{Molloy1995, Albert2000, PhysRevLett.85.4626, Cohen2001, Tanizawa2005}. 
Recently, studies of network resilience has moved their focus to more realistic scenarios of interdependent networks \cite{Buldyrev2010}, different failure~\cite{Gao2015} and recovery \cite{Shekhtman201628, Bttcher2017} mechanisms.

Although the study of network robustness is mature, the majority of the targeted attack strategies are still based on the heuristic identification of influential nodes \cite{FREEMAN1978215, Kitsak2010, KleinbergHITS, Cohen2001, Chen2008EGP} with no performance guarantees for the optimality of the solution. 
Finding the minimal set of nodes such that their removal maximally fragments the network is called network dismantling problem \cite{Braunstein2016, Zdeborova2016} and belongs to the NP-hard class. Thus no polynomial-time algorithm has been found for it and only recently different state-of-the-art methods were proposed as approximation algorithms \cite{Morone2015Nature, Braunstein2016, Zdeborova2016, Morone2016, Tian2017NatComm} for this task. Although state-of-the-art methods \cite{Morone2015Nature, Braunstein2016, Zdeborova2016, Morone2016, Tian2017NatComm} show promising results for network dismantling, we take one step back and analyze the implicit assumption these network dismantling algorithms have. They make implicit assumption that the cost of removing actions are equivalent for all of the nodes, regardless of their centrality in network. 
However, attacking a central node, e.g., a high degree node in socio-technical systems usually comes with the additional cost when compared to the same action on a low degree node. Therefore, it is more realistic to explicitly assume that the cost of an attack is proportional to the number of the edges this attack strategy will remove.

We investigated different state-of-the-art algorithms and the results show that with respect to this new concept of cost, most state-of-the-art algorithms are very inefficient for their high cost, and in most instances perform even worse than random removal strategy.
To overcome this large cost, we proposed a edge-removal strategy, named Hierarchical Power Iterative Normalized cut (HPI-Ncut) as one of the possible solutions.  
Actually, removing a node is equivalent to removing all edges of that node, and therefore all node removal actions can be reproduced with edge removal strategy but vice versa does not hold.
To partition a network, node removal algorithms always remove all the edges connected to some important nodes. However, it is unnecessary to do this because only some specific edges play a key role both on the importance of the nodes and the connectivity of the network.
In cases when the link removal strategies are possible, 
our results show that the edge removal algorithm we used outperforms all the state-of-the-art targeted node attack algorithms.
Finally, we compared the cost of the proposed edge removal strategy HPI-Ncut with other two edge removal strategies which are based on edge betweenness centrality \cite{FREEMAN1978215} and bridgeness centrality \cite{Cheng2010Bridge}.

\section{Results}


A lot of algorithms have been proposed to address network fragmentation problem~\cite{Albert2000,Cohen2001,Chen2008EGP,Altarelli2014,Morone2015Nature} from the node removal perspective. 
These algorithms mainly pay attention to the minimization of the size of the giant connected component and assume that the cost is proportional to the number of removed nodes.
However, the essence of the node removing is to remove all the edges connected to it.
In this article, we make explicit assumption that the cost of removing a node is proportional to the number of the associated edges that has to be removed. This implies that the nodes with higher degree have higher associated removal cost.

In subsection \ref{Data_sets}, we introduce the empirical and artificial networks that are used in this paper.
Then in subsection \ref{cost-auc}, we quantify the cost of the state-of-the-art node removal strategies and show that in most cases the cost of such attacks are inefficient. This results have important impact for real world scenarios of network fragmentations where cost budget is limited.
Finally, when it is possible to remove single edges (e.g. shielding a communication links, removing power lines, cutting off trading relationships or others), we use a spectral edge removal method and compare its cost with other strategies in subsections \ref{edge-removal_res}, \ref{Result_gcc}. 
The effect of edge removal as an immunization measure for the spreading process is shown in subsection \ref{Result_Spread}.

\subsection{Data sets} \label{Data_sets}

To evaluate the performances of the network dismantling (fragmentation) algorithms, both real networks and synthetic networks are used in this paper: (i) \textit{Political Blogs}\cite{Adamic2005} is an undirected social network which was collected around the time of the US. presidential election in 2004. This network is a relatively dense network whose average degree is $27.36$. (ii) \textit{Petster-hamster} is an undirected social network which contains friendships and family links between users of the website \textit{hamsterster.com}. This network data set can be downloaded from KONECT\footnote{\url{http://konect.uni-koblenz.de/networks/petster-hamster}}. (iii) \textit{Powergrid}\cite{Watts1998} is an undirected power grid network in which a node is either a generator, a transformator or a substation, while a link represents a transmission line. This network data set can also be downloaded from KONECT\footnote{\url{http://konect.uni-koblenz.de/networks/opsahl-powergrid}}. (iv) \textit{Autonomous Systems} is an undirected network from the University of Oregon Route Views Project~\cite{leskovec2005}. This network data set can be downloaded from SNAP\footnote{\url{https://snap.stanford.edu/data/as.html}}. (v) Erdős–Rényi (ER) network\cite{erdds1959random} is constructed with $2500$ nodes. Its average degree is 20 and the replacement probability is 0.01. (vi) Scale-free (SF) network with size 10,000, exponent 2.5, and average degree 4.68. (vii) Scale-free (SF) network with size 10,000, exponent 3.5, and average degree 2.35. (viii) Stochastic block model (SBM) with ten clusters is an undirected network with 4232 nodes and average degree 2.60. The basic properties of these networks are listed in table~\ref{Table_property}.

\subsection{Cost-fragmentation inefficiency of the node targeting attack strategies} \label{cost-auc}
Let us define the function $f_{\mathcal{D}}(x)$ as the size of GCC for fixed attack cost $x$ for strategy $\mathcal{D}$. The \textit{cost} $x \in [0,1]$ is measured as the ratio of the number of removed edges in the network. 
Now, for the fixed budget $x$, the strategy $\mathcal{D}$ is more efficient than strategy $\mathcal{L}$ if and only if $f_\mathcal{D}(x) < f_\mathcal{L}(x)$, i.e., size of the GCC is smaller by attacking with strategy $\mathcal{D}$ than with strategy $\mathcal{L}$ with limited budget $x$. 

\begin{figure}
\centering
\includegraphics[width=18cm]{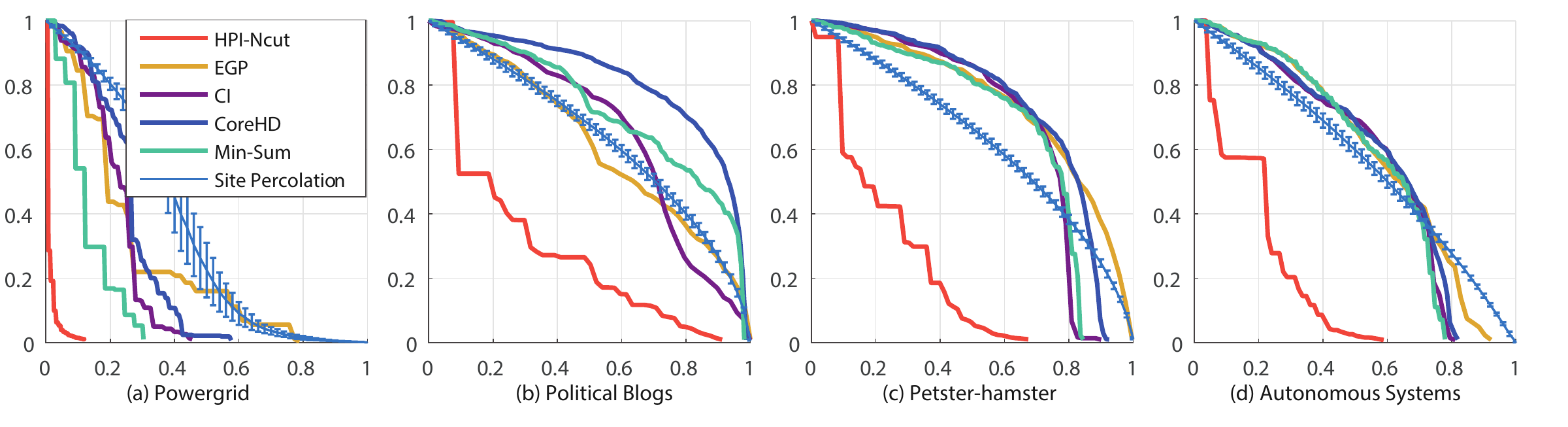}\\
\caption{The size of the GCC of the networks versus the link removing proportion, comparing with classical node removal-based methods on real networks. The results of the site percolation are obtained after 100 independent runs.}\label{fig_gcc_real_node}
\end{figure}

One way to compare the attack performances of strategies is to plot the function $f_{\mathcal{D}}(x)$ of the size of GCC after attack versus the cost, see fig.~\ref{fig_gcc_real_node}. Here we define the \textit{cost-fragmentation effectiveness (CFE)} for strategy $\mathcal{D}$ as the area under the curve of the size of GCC versus the cost, which can be computed as the integral over all possible budgets: $F_\mathcal{D} = \int_0^1 f_\mathcal{D}(x) dx.$
The smaller the CFE (i.e., the area under the curve), the better the attack effect.

\begin{figure}
\centering
\includegraphics[width=18cm]{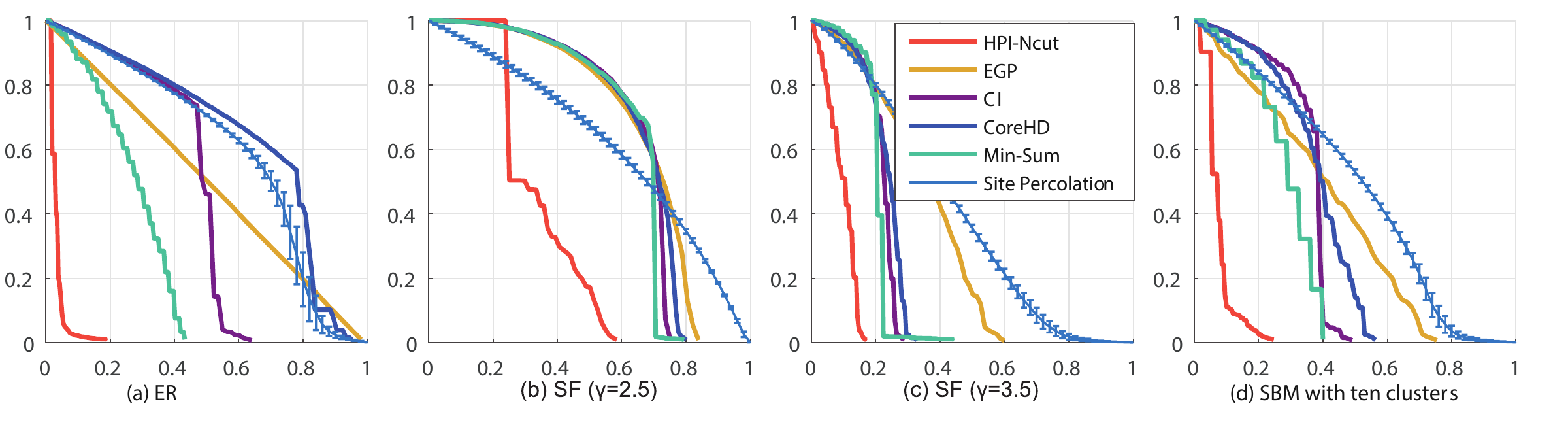}\\
\caption{The size of the GCC of the networks versus link removing proportion, comparing with classical node removal-based methods on artificial networks. The results of the site percolation are obtained after 100 independent runs.}\label{fig_gcc_model_node}
\end{figure}

Taking into account the role of the cost in targeted attacks, the results are highly counterintuitive: For a fixed budget, many networks are more fragile with the High Degree (HD) attack strategy than by High Degree Adaptive (HDA) strategy, as the results shown in table~\ref{Table_Decompose} and table~\ref{Table_Decompose_Improve}.
Furthermore, the performances of the state-of-the-art node removal-based methods can become even worse than the naive random removal of nodes (site percolation) when we take into account the attack cost, as shown in fig.~\ref{fig_gcc_real_node} and fig.~\ref{fig_gcc_model_node}.
In addition, comparing with the CFE of site percolation and bond percolation method in table \ref{Table_Decompose}, we can find that the bond percolation works better on the networks with lower average degree, i.e., on the Powergrid, SF ($\gamma=3.5$), and SBM network, otherwise, it is better to choose the site percolation method.

In fact, networks have their intrinsic resilience under attacking for their distinct network structures. To avoid the interference of the architectural difference of networks, we use site percolation method as a baseline null model. The site percolation strategy randomly removes nodes in a network, which could be used to reflected the intrinsic resilience of the attacked network to a certain extent. The cost-fragmentation effectiveness of the site percolation is denoted with $F_{*} = \int_0^1 f_{*}(x) dx$.\\

Table \ref{Table_Decompose_Improve} summaries the improvement of CFE of different attack strategies $\mathcal{D}$ comparing with the null model (site percolation), which is calculated as $\int_0^1 (f_{*}(x) - f_\mathcal{D}(x)) dx$. On the whole, all node-centric strategies (HD\cite{Cohen2010}, HDA\cite{Cohen2010}, EGP\cite{Chen2008EGP}, CI\cite{Morone2015Nature}, CoreHD\cite{Braunstein2016} and Min-sum\cite{Zdeborova2016}) distinctly work better than baseline on the three networks with lower average degree, i.e., powergrid, SF ($\gamma=3.5$), and SBM network.
However, on empirical social Petster-hamster network, Political Blogs network, Autonomous Systems network and SF ($\gamma=2.5$) network, all node-centric strategies (HD\cite{Cohen2010}, HDA\cite{Cohen2010}, EGP\cite{Chen2008EGP}, CI\cite{Morone2015Nature}, CoreHD\cite{Braunstein2016} and Min-sum\cite{Zdeborova2016})  
are comparably equal or even less cost-fragmentation inefficient than the baseline random model, according to the CFE score. The last line of the table \ref{Table_Decompose_Improve}, the average value of the improvement over different networks is computed, which can reflect the overall CFE of the algorithms. 
This results suggest that state-of-the-art node-centric algorithms in realistic settings are rather inefficient if the cost of fragmentation is taken into account. 

\begin{table}
   \centering
   \caption{Basic statistical features of the GCCs of the eight real and synthetic networks.}
   \label{Table_property}
   \begin{tabular}{l rrcl }
  \hline
  Network & Nodes & Links & Avg. Degree & Sparsity  \\ \hline
  Political Blogs & 1222 & 16714 & 27.36 & 2.24E-2  \\
  Petster-hamster & 2000 & 16098 & 16.10 & 8.05E-3  \\
  Powergrid & 4941  & 6594 & 2.67 & 5.40E-4  \\
  Autonomous Systems & 6474 & 12572 & 3.88 & 6.00E-4 \\
  ER  & 2500  & 12500 & 10.00  & 4.00E-3   \\
  SF ($\gamma=2.5$)  & 10000  & 23423 & 4.68 & 4.69E-4  \\
  SF ($\gamma=3.5$)  & 10000  & 11761 & 2.35  & 2.35E-4 \\
  SBM & 4232  & 5503 & 2.60 & 6.15E-4  \\
  \hline
\end{tabular}
\end{table}

   
\begin{table}
   \centering
   \caption{CFE, i.e., the area under the curve of the size of the GCC after attacking by different algorithms. P$_{site}$ is short for site percolation, P$_{bond}$ for bond percolation, Betw for betweenness, Bridg for bridgeness. The best performing algorithm in each column is emphasized in bold. }\label{Table_Decompose}
    \begin{tabular}{l cccc cccc ccc}
  \hline
  CFE &P$_{site}$ &HD &HDA &EGP &CI &CoreHD &Min-Sum &P$_{bond}$ &Betw &Bridg  & HPI-Ncut  \\ \hline
  Political Blogs &0.638 &0.920 &0.861 &0.619 &0.657 &0.815 &0.726 &0.843 &0.597 &0.910  &\textbf{0.278} \\
  Petster-hamster &0.627 &0.677 &0.696 &0.747 &0.687 &0.736 &0.675 &0.817 &0.536 &0.689  &\textbf{0.224} \\
  Powergrid &0.371 &0.260 &0.293 &0.263 &0.219 &0.256 &0.130 &0.305 &0.145 &0.420 &\textbf{0.014} \\
  Autonomous Systems &0.567 &0.576 &0.604 &0.592 & 0.567 &0.576 &0.567 &0.605 &0.527 &0.618 &\textbf{0.192} \\
  ER  &0.601 &0.547 &0.647 &0.502 &0.441 &0.647 &0.268 &0.753 &0.387 &0.542 &\textbf{0.032} \\
  SF ($\gamma=2.5$) &0.619 &0.700 &0.706 &0.671 &0.650 &0.660 &0.636 &0.683 &0.672 &0.694 &\textbf{0.342} \\
  SF ($\gamma=3.5$) &0.406 &0.231 &0.228 &0.343 &0.214 &0.227 &0.202 &0.298 &0.312 &0.352 &\textbf{0.092} \\
  SBM &0.487 &0.419 &0.378 &0.397 &0.348 &0.374 &0.284 &0.384 &0.348 &0.512 &\textbf{0.075} \\
  \hline
\end{tabular}
\end{table}

\begin{table}
   \centering
   \caption{The improvement of the CFE of each algorithm, comparing with the baseline, i.e., site percolation method. The best performing algorithm in each column is emphasized in bold.}\label{Table_Decompose_Improve}
    \begin{tabular}{l cccc cccc cc}
  \hline
  Improvement &P$_{bond}$ &HD &HDA &EGP &CI &CoreHD &Min-Sum &Betw &Bridg & HPI-Ncut \\ \hline
  Political Blogs &-32\% &-44\% &-35\% &3\% &-3\% &-28\% &-14\% &6\% &-43\% &\textbf{56\%} \\
  Petster-hamster &-30\% &-8\% &-11\% &-19\% &-10\% &-17\% &-8\% &15\% &-10\% &\textbf{64\%} \\
  Powergrid &18\% &30\% &21\% &29\% &41\% &31\% &65\% &61\% &-13\% &\textbf{96\%} \\
  Autonomous Systems &-7\% &-2\% &-7\% &-4\% &0 &-2\% &0\% &7\% &-9\% &\textbf{66\%}      \\
  ER  &-25\% &9\% &-8\% &17\% &27\% &-8\% &55\% &36\% &10\% &\textbf{95\%}  \\
  SF ($\gamma=2.5$) &-10\% &-13\% &-14\% &-8\% &-5\% &-7\% &-3\% &-9\% &-12\% &\textbf{45\%} \\
  SF ($\gamma=3.5$) &27\% &43\% &44\% &15\% &47\% &44\% &50\% &23\% &13\%  &\textbf{77\%}  \\
  SBM &20\% &13\% &22\% &18\% &28\% &23\% &41\% &28\% &-6\%  &\textbf{84\%}  \\  \hline
  Average &-5\% &3\% &2\% &6\% &16\% &5\% &23\% &21\% &-9\%  &\textbf{73\%}  \\  \hline
\end{tabular}
\end{table}

\subsection{The edge-removal problem} 
\label{edge-removal_res}


In network science, nodes represent entities in a system and edges represent the relationships or interactions between them. Both the nodes and the edges are the fundamental part of a network. Deleting or removing a specific ratio of them will lead to great changes in the structure and functions of the network. The problem of network attack or fragmentation has received a huge amount of attention in the past decade~\cite{Pastor2002,Chen2008EGP,Pastor2015,Zhang2016,Wang20161}. However, as far as we are concerned, almost all of the attack strategies are node removal based, in which the node removal operation is carried out via removing all the edges connected to it. In fact, to partition a network in to small clusters, it is unnecessary to remove all the links of a node. We remove a node because either we suppose it has a higher influence or the node is a bridge between clusters. If we remove part of its connected links, its influence may greatly reduce or it may not be a bridge any more. 
From another perspective, edges play far different roles in real networks~\cite{Binder2012,bakshy2012}. Some of them are crucial to the diffusion process, while others are irrelevant. Thus, if the edge removal actions on networks are applicable, the edge removal attack will be more accuracy and efficient. 


The link fragmentation or attack problem can be narrated as follows: If we have a budget of $x$ links that can be attacked or removed, which links should we pick? This is mathematically equivalent to asking how to partition a given network with a minimal separate set of edges. The objective function of link attack takes the following general form~\cite{VonLuxburg2007}:
\begin{equation}
 cut(A_1,\cdots, A_k) := \frac{1}{2} \sum_{i=1}^{k}{W(A_i,\overline{A_i})} 
\end{equation}
where $A_1,\cdots, A_k$ are $k$ nonempty subsets from a partition of the original network, $\overline{A_i}$ is the complement set of the nodes of $A_i$, and $W(A_i,\overline{A_i})$ is the number of the links between the two disjoint subsets $A_i$ and $\overline{A_i}$.

In this paper, we applied the spectral strategy for edge attack problem, which fall in the class of well known spectral clustering and partitioning algorithms \cite{Bisestion1991, shi2000Ncut, SC, SGT, NewmanSC}. We use the hierarchical partitioning with Ncut objective function\cite{shi2000Ncut} combined with power iteration procedure for approximation of eigenvectors. The complete description of this HPI-Ncut edge removal strategy will be presented in the Section~\ref{methods}. The results show that the HPI-Ncut strategy greatly decreases the cost of the attack, comparing with the state-of-the-art nodes removing strategies. 
In the following subsection, we will compare HPI-Ncut algorithm with random uniform attack strategy, edge betweenness, bridgeness, and some classical node removing strategies (see the definitions of these algorithms in the Section~\ref{methods}).

\subsection{Effectiveness of the HPI-Ncut algorithm} \label{Result_gcc}

\begin{figure}
  \centering
  \includegraphics[width=18cm]{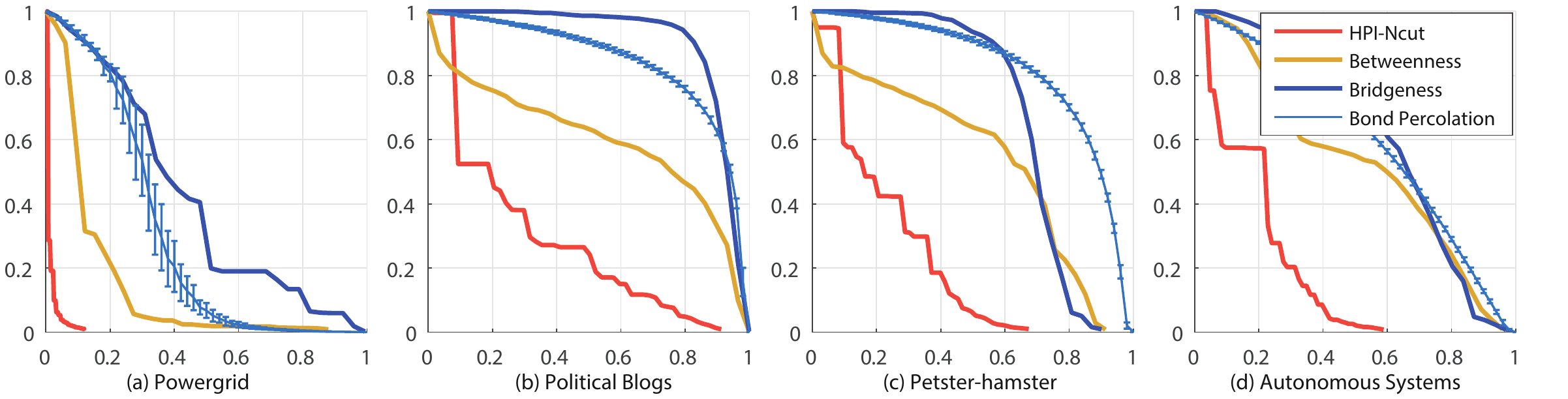}\\
  \caption{The size of the GCC of the networks versus link removing proportion, comparing with existed link removal-based methods on real networks. The results of the bond percolation are obtained after 100 independent runs.}\label{fig_gcc_real_link}
\end{figure}

\begin{figure}
  \centering
  \includegraphics[width=18cm]{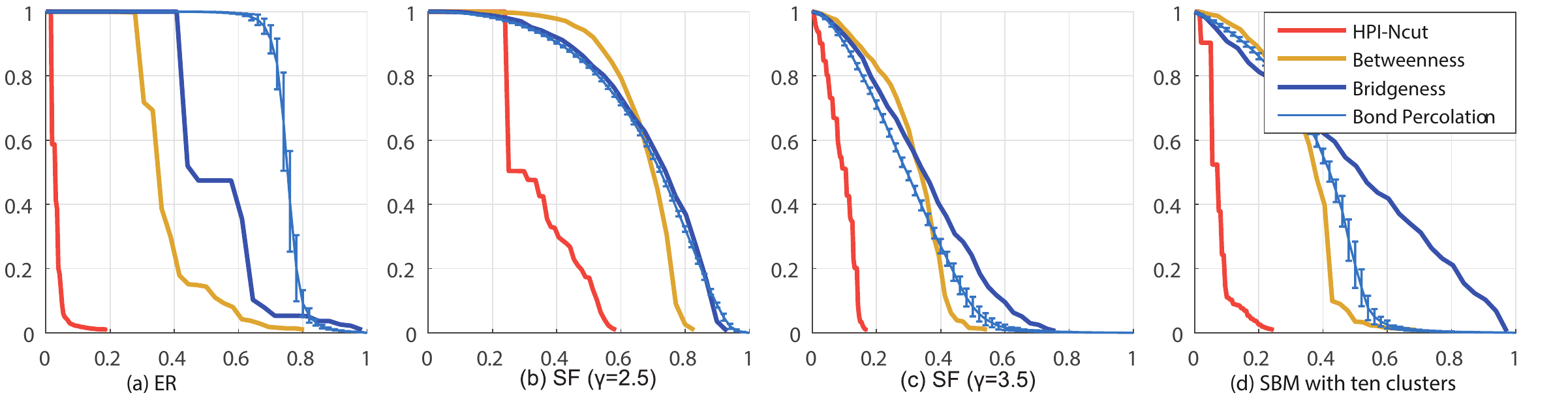}\\
  \caption{The size of the GCC of the networks versus link removing proportion, comparing with existed link removal-based methods on artificial networks. The results of the bond percolation are obtained after 100 independent runs.}\label{fig_gcc_model_link}
\end{figure}

In general case, each attack strategy algorithm could generate a ranking list of all (or partial) nodes or links of the network. After removing the nodes or links one after another, the size of the GCC of the residual network characterizes the effectiveness of each algorithm. The removal process will cease when the size of the GCC is smaller than a given threshold (here we use 0.01).
In this paper, to test the effectiveness of this spectral edge removal algorithm, HPI-Ncut, we plot the size of the GCC versus the removal fraction of links, for both real networks (fig.~\ref{fig_gcc_real_node} and fig.~\ref{fig_gcc_real_link}) and synthetic networks (fig.~\ref{fig_gcc_model_node} and fig.~\ref{fig_gcc_model_link}), comparing with classical node removing algorithms (fig.~\ref{fig_gcc_real_node} and fig.~\ref{fig_gcc_model_node})  and existed link evaluation methods (fig.~\ref{fig_gcc_real_link} and fig.~\ref{fig_gcc_model_link}). The results show that the HPI-Ncut algorithm outperforms all the other attack algorithms.

In fig.~\ref{fig_gcc_real_node} and fig.~\ref{fig_gcc_model_node}, we compared the HPI-Ncut algorithm with some state-of-the-art node removal-based target attack algorithms.
Fig.~\ref{fig_gcc_real_node} (a) shows that all the node removal-based algorithms are better than the site percolation method on Powergrid network, that is because the average degree of the Powergrid network is very low, only 2.67. This could also be verified by the results in fig.~\ref{fig_gcc_model_node} (c) and (d), in which the average degree of the SF ($\gamma=3.5$) and the SBM network are 2.35 and 2.60, respectively. The trends of the curves in fig.~\ref{fig_gcc_real_node} and fig.~\ref{fig_gcc_model_node} also show that the target attack algorithms works better on networks with lower average degree. 
Furthermore, regardless of the HPI-Ncut algorithm, other algorithms have poorer performance than baseline method (site percolation). The performances of site percolation are better until the proportion of the removed links is greater than 0.7 on SF ($\gamma=3.5$) network and until the proportion is greater than 0.2 on SF ($\gamma=2.5$) network.
The site percolation on the SF ($\gamma=3.5$) presents an obvious phase transition phenomenon \cite{Cohen2010} comparing with the result on the SF ($\gamma=2.5$).
In addition, in fig.~\ref{fig_gcc_model_node} (a) and (d), the SBM network has obvious clusters structure comparing with the ER network. The Min-Sum, CI, CoreHD, EGP, and site percolation algorithms have a better performance on the SBM network. Moreover, the error of the site percolation method on the ER network is larger than the error on SBM network. That implies that the cluster structure of a network has a big influence on the performance of the attack strategies.

To conclude the results of fig.~\ref{fig_gcc_real_node} and fig.~\ref{fig_gcc_model_node}, the state-of-the-art targeted node removal strategies make large cost for optimized targeted attacks. 
Contrary, HPI-Ncut algorithm overwhelmingly outperforms all the node removal-based attack algorithms, no matter on sparse or dense networks, on the networks with or without clusters structure.

\begin{figure}
  \centering
  \includegraphics[width=18cm]{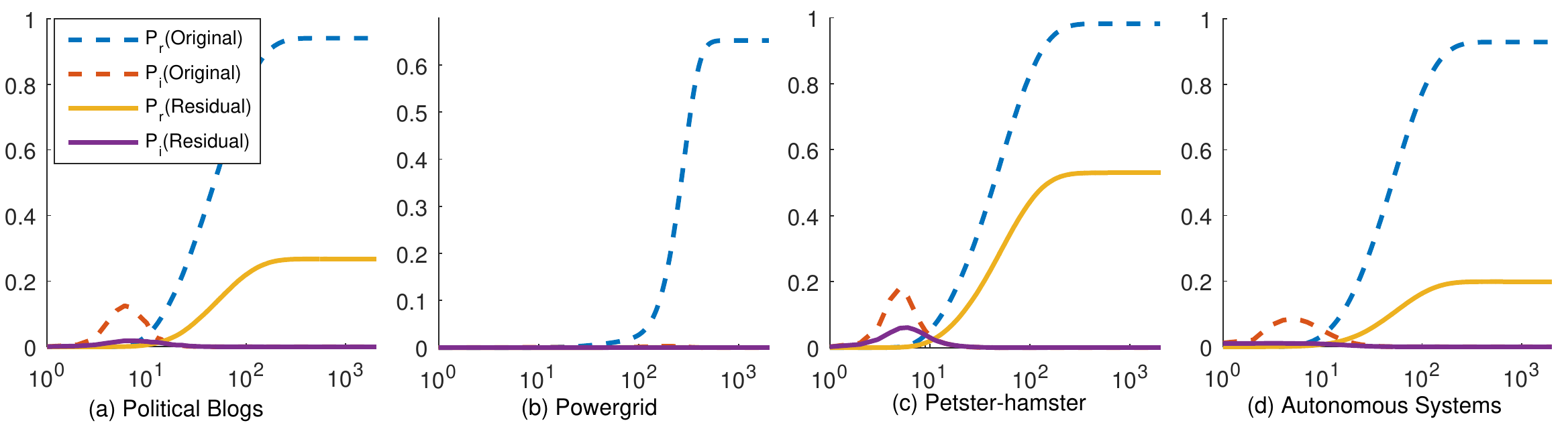}\\
  \caption{The spreadabilities of the networks before and after the removing of 10\% edges by HPI-Ncut algorithm. The x-axis is the time unites. $P_i$ is the number of infected entities and $P_r$ is the number of recovered entities in the network. In the SIR model, the infection rate $\beta$ is 0.10, the recovery rate is 0.02, and the basic reproduction number is 5. All the results are the average of 100 times independent runs. It is worth noting that the size of GCC of the Powergrid network is only 54 after removing 10\% of links by HPI-Ncut algorithm. }\label{fig_SIR}
\end{figure}

In fig.~\ref{fig_gcc_real_link} and fig.~\ref{fig_gcc_model_link}, we compared the HPI-Ncut algorithm with some exited link evaluation algorithms. First of all, we can find that the HPI-Ncut algorithm works better and is more stable than all the other algorithms.
Secondly, comparing with the results of site and bond percolation in fig.~\ref{fig_gcc_real_node} and fig.~\ref{fig_gcc_model_node}, we can see that the bond percolation method outperforms the site percolation method only when the average degree of the network is lower (see the results of the Powergrid, SF ($\gamma=3.5$), and SBM network), otherwise, the site percolation is a better choice. 
Thirdly, in the fig.~\ref{fig_gcc_model_link} (b) and (c), we can see that the bond percolation method have a better performance comparing with the edge betweenness and bridegeness algorithm when the cost is limited on scale-free networks, i.e., the proportion of the removed links is smaller than 0.63 in fig.~\ref{fig_gcc_model_link}(b) and is smaller than 0.4 in fig.~\ref{fig_gcc_model_link}(c).



To conclude, the HPI-Ncut algorithm overwhelmingly outperforms all the node removal-based attack algorithms and link evaluation algorithms, no matter on sparse or dense networks, on networks with or without clusters structure.

\subsection{Spreading dynamics after spectral edge immunization/attack} \label{Result_Spread}
To more intuitively display the targeted attack by HPI-Ncut, we studied the susceptible-infectious-recovery (SIR) ~\cite{Hethcote2000} epidemic spreading process on four real networks. We compared both the spreading speed and spreading scope on these networks before and after targeted immunization by HPI-Ncut. The simulation results in fig.~\ref{fig_SIR} show that, by simply removing 10\% of links, the function of the networks had been profoundly affected by the HPI-Ncut immunization. The proportion of the GCC of the Political Blogs, Powergrid, Petster-hamster, and Autonomous Systems network after attack are 37\% (449/1222), 1\% (54/4941), 57\% (1146/2000), and 37\% (2387/6474), respectively. Thus, the spreading speeds are greatly delayed and the spreading scoops are tremendously shrunken on these networks.

\section{Methods}\label{methods}

\subsection{Some existed attack strategies}
In this subsection, we will briefly introduce some state-of-the-art node removal attack algorithms and some edge evaluation methods used in this paper. The two edge evaluation methods, i.e., edge betweenness and bridgesness, are used to measure the importance or significance of links in spread dynamics or structure connectivities of networks. We use them as comparable link removal-based attack algorithms in this paper.

\begin{itemize}
\item Percolation method. In percolation theory\cite{Callaway2000}, node of networks usually called `site', while edge usually called `bond'. In the study of the network attack, percolation is a random uniform attack method which either removes nodes randomly (site percolation) or removes edges randomly (bond percolation).

\item Equal graph partitioning (EGP) algorithm. EGP algorithm\cite{Chen2008EGP}, which is based on the nested dissection \cite{Lipton1979} algorithm, can partition a network into two groups with arbitrary size ratio. In every iteration, EGP algorithm divides the target nodes set into three subsets: first group, second group, and the separate group. The separate group is made up of all the nodes that connect to both the first group and the second group. Then minimize the separate group by trying to move nodes to the first group or the second group. Finally, after removing all the nodes in the separate group, the original network will be decomposed into two groups. In our implementation, we partition the network into two groups with approximate equal size. 

\item Collective Influence (CI) algorithm. CI algorithm\cite{Morone2015Nature} attacks the network by mapping the integrity of a tree-like random network into optimal percolation theory~\cite{kovacs2015network} to identify the minimal separate set. Specifically, the collective influence of a node is computed by the degree of the the neighbors belonging to the frontier of a ball with radius $l$. CI is an adaptive algorithm which iteratively removes the node with highest CI value after computing the CI values of all the nodes in the residual network. In our implementation, we compute the CI values with $l=3$.

\item Min-Sum algorithm. The three-stage Min-Sum algorithm~\cite{Braunstein2016} includes: (1) Breaking all the circles, which could be detected form the 2-core\cite{Kitsak2010} of a network, by the Min-Sum message passing algorithm, (2) Breaking all the trees larger than a threshold $C_1$, (3) Greedily reinserting short cycles that no greater than a threshold $C_2$, which ensures that the size of the GCC is not too large. In our implementation, we set $C_1$ and $C_2$ as 0.5\% and 1\% of the size of the networks.

\item CoreHD algorithm. Inspired by Min-Sum algorithm,
CoreHD algorithm\cite{Zdeborova2016} iteratively deletes the node with highest degree from the the 2-core\cite{Kitsak2010} of the residual network.

\item Edge betweenness centrality\cite{Freeman1977}. Betweenness is a widely used centrality measure which is the sum of the fraction of all-pairs shortest paths that pass a node. Edge betweenness, an extension of the betweenness, is used to evaluate the importance of a link, and is defined as the sum of the fraction of all-pairs shortest paths that pass this link\cite{Lu2016}. 

\item Bridgeness\cite{Cheng2010Bridge}. Bridgeness use local information of the network topology to evaluate the significance of edges in maintaining network connectivity. The bridgeness of a link is determined by the size of $k$-clique communities that the two end points of this link are connected with and the size of the $k$-clique communities that this link is belonging to.

\end{itemize}

\subsection{Hierarchical Power Iterative Normalized cut (HPI-Ncut) edge removal strategy }
Here we describe the hierarchical iterative algorithm for edge removal strategy. This algorithm hierarchically applies the spectral bisection algorithm, which has the same objective function as the Normalized cut algorithm\cite{shi2000Ncut}. Furthermore we have used the power iteration method to approximate spectral bisection. We provide proof on the exponential convergence and asymptotic upper bounds for the run-time complexity.\\ \\
In order to explain our algorithm, we quickly recall the spectral bisection algorithm.\\ \\
\textbf{The spectral bisection algorithm}\\
Input: Adjacency matrix $W$ of a network\\ 
Output: A separated set of edges that partition the network into two disconnected clusters $A$, $\bar{A}$.
\begin{enumerate}
\item Compute the eigenvector $v_2$, which corresponds to the second smallest eigenvalue of the normalized Laplacian matrix $L_w= D^{-\frac{1}{2}}\left(D-W \right) D^{-\frac{1}{2}}$, or some other vector $v$ for which $\frac{v^TL_w v}{v^Tv}$ is close to minimal. We use the power iteration method to compute this vector, which will be explained later.
\item Put all the nodes with $v_2(i)>0$ into the first cluster $A$ and all the nodes with $v_2(i)\leq 0$ into the second cluster $\bar{A}$. All the edges between these two clusters form the separation set that can partition the network.
\end{enumerate}

The clusters that we obtained by this method had usually very balanced sizes. If, however, it is very important to get clusters of exactly the same size, one could put those $\frac{n}{2}$ nodes with the largest entries in $v_2$ into one cluster and the remaining nodes into the other cluster.\\ \\
\textbf{Hierarchical Power Iterative Normalized cut (HPI-Ncut) algorithm}\\
Input: Adjacency matrix of a network\\
Output: Partition of the network into small groups
\begin{enumerate}
\item Partition the GCC of the network into two disconnected clusters $A$ and $\bar{A}$ by using the spectral bisection algorithm and removing all the links in the separated set.
\item If the budget for link removal has not been overrun, and if the GCC is not yet small enough, partition $A$ and $\bar{A}$ with step 1, respectively.
\end{enumerate}

The reason why we cluster hierarchically is that, this allows us to refine the fragmentation gradually.
For example, if after partitioning the network into $2^k$ clusters, we decide that the clusters should be smaller, we would just have to partition each of the existing clusters into $2$ new clusters, obtaining $2^{k+1}$ clusters. So the links that were attacked already remain attacked and we just need to attack some additional ones. If, however, we had used spectral clustering straightforwardly, it could happen that the set of links to be attacked in order to partition the network into $2^{k+1}$ clusters, would not contain the set of links that needed to be attacked for $2^k$ clusters.\\ \\
\textbf{Power iteration method}\label{powerMethod}\\
Input: Adjacency matrix $W$ of a network\\
Output: The eigenvector $v_2$ or some other vector $v$ for which $\frac{v^TL_wv}{v^Tv}$ is close to $\lambda_2$.
\begin{enumerate}
\item Draw $v$ randomly with uniform distribution on the unit sphere.
\item Set $v\leftarrow v-v_1^Tv\cdot v_1$, where $v_1=\frac{1}{\sqrt{n}}(1,...,1)$.
\item For $i=1$ to $\eta (n)$ { \\
$v\leftarrow \frac{\tilde{L} v}{\Vert \tilde{L} v\Vert } $, where $\tilde{L}=2\cdot I-L_w$ and $L_w= D^{-\frac{1}{2}}\left(D-W \right)   D^{-\frac{1}{2}}$.\\
}
\end{enumerate}

\subsection*{Objective function of the spectral bisection algorithm}
In appendix A, we show that the spectral bisection algorithm has the same objective function with the relaxed Ncut \cite{shi2000Ncut} algorithm:
\begin{equation}
Ncut(A, \bar{A}):= \sum_{i\in A, j \in \bar{A}} W_{i,j} \left(\frac{1}{\sum_{i\in A} D_{ii}}+\frac{1}{\sum_{j\in \bar{A}} D_{jj}} \right),
\end{equation}
where $A\subseteq V$ denotes set of nodes in the first partition, $\bar{A}$ the set of nodes in the second partition and $D_{ii}$ is the degree of the node $i$.

The main reason we used this objective function is that it minimizes the number of links that are removed and the total sum of node degree centralities in both partition $A$ and $\bar{A}$ is approximately equal. 

In appendix B, we show the exponential convergence of the power iteration method to the eigenvector associated with the second smallest eigenvalue of $L_w$. 

\subsection*{Complexity}
In appendix C, we show that the complexity of the spectral bisection algorithm is $O(\eta (n)\cdot n\cdot \bar{d})$ and the complexity of the hierarchical clustering algorithm is $O(\eta (n)\cdot n\cdot \bar{d}\cdot \log(n))$ where $\eta (n)$ is the number of iterations in the power iteration method. The power iteration method converges with exponential speed as $\eta (n) \to \infty$. The average degree $\bar{d}$ is almost constant for large sparse network. Hence we may expect assymptotically good results with $\eta (n)=\log (n)^{1+\epsilon}$ for any $\epsilon  >0$, giving the hierarchical spectral clustering algorithm a complexity of $O( n \cdot \log^{2+\epsilon}(n))$. In practice, we have used $\epsilon=0.1$, which gives a complexity of $O( n \cdot \log^{2.1}(n))$.

\section{Conclusion}
To summarize, we investigated some state-of-the-art node target attack algorithms and found that they are very inefficient when the cost of the attack is taken into consideration. The cost of removing a node is defined as the number of links that are removed in the attack process. 
We found some highly counterintuitive results, that is, the performances of the state-of-the-art node removal-based methods are even worse than the naive site percolation method with respect to the limited cost.
This demonstrates that the current state-of-the-art node targeted attack strategies underestimate the heterogeneity of the cost associated to node in complex networks. 

Furthermore, in cases when the link removal strategies are possible, we compared the performances of the node-centric (HD\cite{Cohen2010}, HDA\cite{Cohen2010}, EGP\cite{Chen2008EGP}, CI\cite{Morone2015Nature}, CoreHD\cite{Braunstein2016} and Min-sum\cite{Zdeborova2016}) and edge removal strategies (edge betweenness \cite{Freeman1977} and bridgeness\cite{Cheng2010Bridge} strategy) based on the cost of their attacks, which are measured in the same units, i.e., the ratio of the removed links. Node removal-based algorithms always deletes all the links respected to the removed nodes which is not economical respect to the limited cost. In order to resolve the high-cost problem in network attack,  we proposed a hierarchical power iterative algorithm (HPI-Ncut) to partition networks into small groups via edge removing, which has the same objective function with the Ncut \cite{shi2000Ncut} spectral clustering algorithm. The results show that HPI-Ncut algorithm outperforms all the node removal-based attack algorithms and link evaluation algorithms on all the networks. In addition, the total complexity of the HPI-Ncut algorithm is only $O( n \cdot \log^{2+\epsilon}(n))$.

\section*{Acknowledgements}
The work of N.A.-F. has been funded by the EU Horizon 2020 SoBigData project under grant agreement No. 654024.
The work of D.T. is funded by the by the Croatian Science Foundation IP-2013-11-9623 "Machine learning algorithms for insightful analysis of complex data structures". X.-L.R. thanks the support from China Scholarship Council (CSC).



\bibliography{sample.bib}

\newpage

\section*{Appendix A: Objective function}
Let $G=(V, E)$ be an undirected graph with adjacency matrix $W$ and diagonal degree matrix $D$, whose $i$-th entry $ D_{ii}=\sum_{j=1}^n W_{ij}$, is the degree of the node $i$. 
For $A\subseteq V$, let $cut(A)$ denote the number of links between $A$ and its complement $\bar{A}$. We define
\begin{equation}
Ncut(A, \bar{A}):=cut(A, \bar{A})\left(\frac{1}{assoc(A)}+\frac{1}{assoc(\bar{A})} \right). 
\end{equation}
where $assoc(A)=\sum_{i\in A} D_{ii}$. 
If we describe the set $A$ by the normalized indicator vector \begin{equation}
x_A(i):=
\begin{cases}
1 & \ if \ i \in \ A, \\
-\frac{\sum_{j\in A} D_{jj}}{\sum_{j\in B} D_{jj}} & \ otherwise
\end{cases}
\end{equation}
one can show\cite{shi2000Ncut} that\begin{equation}\label{iden}\min_{A\subseteq V}
Ncut(A, \bar{A})=\min_{\substack{A\subseteq V, \\  }} \frac{x_A^T\left(D-W \right) x_A}{x_A^TDx_A}.
\end{equation}
From the definition of $Ncut$ one can see that finding a set $A$ which minimizes $Ncut(A, \bar{A})$ corresponds to partitioning the network into two sets $A$ and $\bar{A}$ such that \begin{enumerate}
\item $cut(A, \bar{A})$ is small and hence there are only few links between $A$ and $\bar{A}$
\item $\left(\frac{1}{assoc(A)}+\frac{1}{assoc(\bar{A})} \right)$ is small and so the sets $A$ and $\bar{A}$ contain more or less equally many links.
\end{enumerate}
Finding such a set $A$ is NP-hard ~\cite{shi2000Ncut}, but by relaxing the constraints in the RHS of the identity (\ref{iden}) one can find good approximate solutions $\tilde{A}$:
\begin{enumerate}
\item Find \begin{equation}\label{Lap}
x_{relaxed}:=arg \min_{\substack{x\in\mathbb{R}^n, x\neq 0 \\ x^TD\vec{1}=0}} \frac{x^T\left(D-W \right) x}{x^TDx},
\end{equation}
 where we have imposed the condition $x^TD\vec{1}=0$, because every set $A$ for which $x_A$ is nontrivial, satisfies $x_A^TD\vec{1}=0$.
\item Set \begin{equation}
 \chi_{\tilde{A}}(i)=round(x_{relaxed}(i)):=\begin{cases}
1 & \ if \ x_{relaxed}(i)\geq 0, \\
-1 & \ otherwise 
\end{cases}
\end{equation} and define $\tilde{A}=\left\lbrace i\in Nodes\vert \chi_{\tilde{A}}(i)=1 \right\rbrace$.
\end{enumerate}
The idea behind this method is that $\chi_{\tilde{A}}$ will be the best approximation of $x_{relaxed}$, out of the set of all vectors with entries in $ -1$ and $1  $, and since $x_{relaxed}$ minimizes 
$
\frac{x^T\left(D-W \right) x}{x^TDx},
$ 

\begin{equation} Ncut(\tilde{A})=\frac{x_{\tilde{A}}^T\left(D-W \right) x_{\tilde{A}}}{x_{\tilde{A}}^TDx_{\tilde{A}}}\approx \frac{\chi_{\tilde{A}}^T\left(D-W \right) \chi_{\tilde{A}}}{\chi_{\tilde{A}}^TD\chi_{\tilde{A}}}\end{equation} will be also close to \begin{equation}
\min_{A\subseteq Nodes} Ncut(A)=\min_{A\subseteq Nodes}\frac{x_A^T\left(D-W \right) x_A}{x_A^TDx_A}.
\end{equation}

One can show that a solution to (\ref{Lap}) is given by $x_{relaxed}=D^{\frac{1}{2}} v_2$, where $v_2$ is the eigenvector of the second smallest eigenvalue $\lambda_2$ of the \textbf{ normalized Laplacian matrix} \begin{equation}
L_w= D^{-\frac{1}{2}}\left(D-W \right)   D^{-\frac{1}{2}}.
\end{equation}
$D$ is a diagonal matrix and if the network is connected we have $D_{ii}>0$.
So the entries of the vectors $x_{relaxed}$ and $v_2$ have the same sign and therefore we have $round(x_{relaxed})=round(v_2)$.

\section*{Appendix B: Exponential convergence of the power iteration method}
$L_w$ is real and symmetric. Therefore it has real eigenvalues $\lambda_1\leq \lambda_2\leq...\leq \lambda_n$ corresponding to eigenvectors $v_1,...,v_n$ which form an orthonormal basis of $\mathbb{R}^n$. One can easily show that $\lambda_1 = 0$ and $\lambda_n\leq 2$. So in order to compute $v_2$ we consider the matrix $\tilde{L}=2\cdot I-L_w$, which has the same eigenvectors $v_1,...,v_n$ as $L_w$. Now the corresponding eigenvalues are $\tilde{\lambda_1}=2\geq...\geq\tilde{\lambda_n}=d_{max}-\lambda_n\geq 0$ and in particular $v_1$ corresponds to the largest eigenvalue and $v_2$ to the second largest eigenvalue.

If $v$ is a random vector uniformly drawn from the unit sphere and we force it to be perpendicular to $v_1$ by setting $v\leftarrow v-v_1^Tv\cdot v_1$, then $v=\psi_2 v_2+...+\psi_n v_n$ and $\psi_2\neq 0$ almost surely. Furthermore $\tilde{L} v=\tilde{\lambda_2}\psi_2 v_2+...+\tilde{\lambda_n}\psi_n v_n $ and if we set $v^{(k)}:=\tilde{L}^k v$, then \begin{equation}\label{conveig}\begin{split}
\frac{v^{(k)}}{\Vert v^{(k)}\Vert}&=\frac{\tilde{\lambda_2}^k\psi_2 v_2+...+\tilde{\lambda_n}^k\psi_n v_n}{\Vert\tilde{\lambda_2}^k\psi_2 v_2+...+\tilde{\lambda_n}^k\psi_n v_n\Vert}\\
&=\frac{\psi_2 v_2+\left(\frac{\tilde{\lambda_3}}{\tilde{\lambda_2}}\right) ^k\psi_3 v_3+...+\left(\frac{\tilde{\lambda_n}}{\tilde{\lambda_2}}\right) ^k\psi_n v_n}{\Vert\psi_2 v_2+\left(\frac{\tilde{\lambda_3}}{\tilde{\lambda_2}}\right) ^k\psi_3 v_3+...+\left(\frac{\tilde{\lambda_n}}{\tilde{\lambda_2}}\right) ^k\psi_n v_n\Vert}
\end{split}
\end{equation}
converges with exponential speed to some eigenvector of $L$ with eigenvalue $\lambda_2$, because for every $i$ with $\lambda_i>\lambda_2$ we have $\frac{\tilde{\lambda_i}}{\tilde{\lambda_2}}<1$ and therefore $\left(\frac{\tilde{\lambda_i}}{\tilde{\lambda_2}}\right) ^k\psi_i v_i\rightarrow 0$. Generally one can deduce from (\ref{conveig}) that
\begin{equation}
\frac{{v^{(k)}}^T L_w v^{(k)}}{{v^{(k)}}^T v^{(k)}}=\frac{\lambda_2\vert \psi_2\vert ^2 +\lambda_3\vert \left(\frac{\tilde{\lambda_3}}{\tilde{\lambda_2}}\right) ^k\psi_3\vert ^2+...+\lambda_n \vert \left(\frac{\tilde{\lambda_n}}{\tilde{\lambda_2}}\right) ^k\psi_n\vert ^2}{|\psi_2|^2 +\vert \left(\frac{\tilde{\lambda_3}}{\tilde{\lambda_2}}\right) ^k\psi_3\vert ^2+...+\vert \left(\frac{\tilde{\lambda_n}}{\tilde{\lambda_2}}\right) ^k\psi_n\vert ^2}
\end{equation} 
and therefore this quantity converges to $\lambda_2$ with exponential speed.

\section*{Appendix C: Complexity}
The complexity of the spectral bisection algorithm is the same as the complexity of the power iteration method. 
The complexity of the power iteration method equals the number of iterations $\eta (n)$ times the complexity of multiplying $\tilde{L}$ and $v$. That is $O(\eta (n)\cdot n\cdot \bar{d})$ where $\bar{d}$ is the average degree of the network, or equivalently $O(|E|\cdot \eta (n))$ where $|E|$ is the number of edges.

Assuming that the spectral bisection algorithm always produces clusters of equal size, the complexity of the hierarchical spectral clustering algorithm is then given by the sum of: \begin{itemize}
\item The complexity of applying spectral bisection once on the whole network. $\rightarrow O(\eta (n)\cdot n\cdot \bar{d})$.
\item The complexity of applying it on each of the two clusters that we obtained from the first application of spectral bisection and which will have size $\frac{n}{2}$. 
\item The complexity of applying it on each of the 4 clusters that we obtained from the previous step and which will have size $\frac{n}{4}$. 
\item The complexity of applying it on each of the $\frac{n}{2}=2^{\log_2(n)-1}$ clusters that we obtained from the previous step and which will have size $\frac{n}{2^{\log_2(n)-1}}=2$. \\
\end{itemize}
That is in total at most \begin{equation}\begin{split}
&O(\eta (n)\cdot n\cdot \bar{d})+2\cdot O(\eta (n)\cdot \frac{n}{2}\cdot \bar{d})+...+2^{\log_2(n)-1}\cdot O(\eta (n)\cdot \frac{n}{2^{\log_2(n)-1}}\cdot \bar{d})\\
&=\sum_{i=0}^{\log_2(n)-1} 2^{i}\cdot O(\eta (n)\cdot \frac{n}{2^{i}}\cdot \bar{d})\\ &= O(\eta (n)\cdot n\cdot \bar{d}) \sum_{i=0}^{\log_2(n)-1} 1\\ &=O(\eta (n)\cdot n\cdot \bar{d}\cdot \log_2(n)),
\end{split}
\end{equation}
where we have made the pessimistic assumption that the number of iterations and the average degrees are in each step as large as they were in the beginning.

The choice of the function $\eta (n)$ is a little bit involved. If the initial random choice of the vector $v$ is very unfortunate, there may be many iterations needed in order to have a good approximation of the eigenvector $v_2$. In fact, if $\psi_2=0$, then this algorithm would not converge to $v_2$ at all, however this event has probability $0$. 

Another condition that might slow down the computation of $v_2$ is if some of the other eigenvalues $\lambda_i$, $i\geq 3$ are close to $\lambda_2$. In that case $\frac{\tilde{\lambda_i}}{\tilde{\lambda_2}}$ would be close to $1$ and therefore one can see from equation (\ref{conveig}) that the correspoding $v_i$ might have a large contribution in $v^{(k)}$ for a long time. However when $\lambda_i$ is close to $\lambda_2$, this also implies that \begin{equation}
\frac{v_i^T L_w v_i}{v_i^T v_i}=\lambda_i
\end{equation}
is close to 
\begin{equation}
\min_{\Vert v\Vert \neq 0} \frac{v^T L_w v}{v^T v}=\lambda_2
\end{equation}
and therefore also provides a good partition of the network, since these are the quantities that are related to the cut-size.


Due to this fast convergence, one can expect assymptotically good partitions when $\eta (n) =\log(n)^{1+\epsilon}$ and $\epsilon >0$,  giving the hierarchical spectral clustering algorithm a complexity of $O( n \cdot \bar{d}\cdot \log^{2+\epsilon}(n))$ in general and $O( n \cdot \log^{2+\epsilon}(n))$ for sparse networks.
 
\section*{Appendix D: HPI-Ncut algorithm with different number of partitions }

\begin{figure}
  \centering
  \includegraphics[width=16cm]{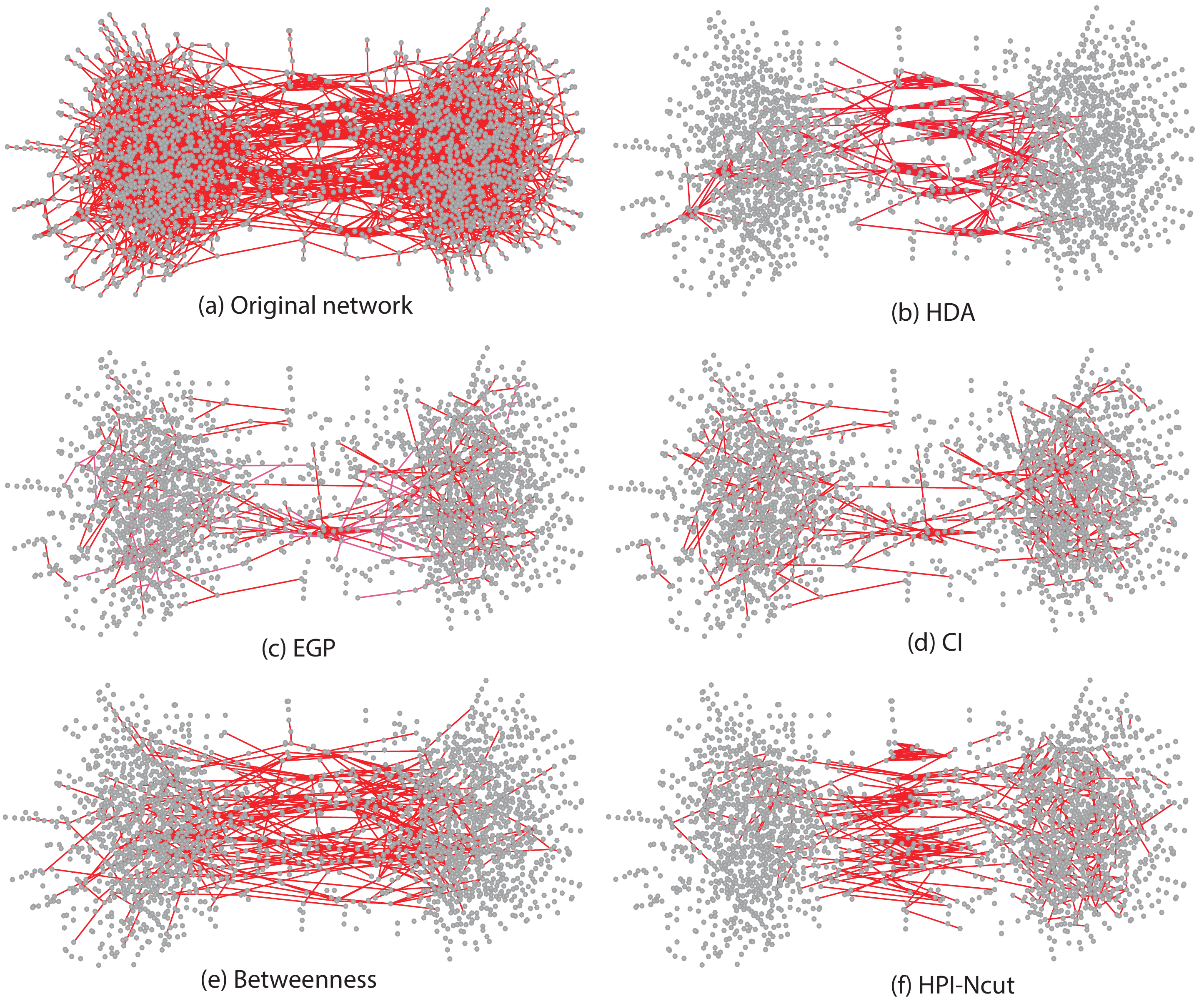}\\
  \caption{The schematic diagram of the removed links in a SBM network with two clusters. (a) is the original network with all the links. (b)-(f) are the top 10\% links (i.e., 373 links) removed by different algorithms.}\label{fig_remove}
\end{figure}

Previous sections give us an clear picture about the performances of different attack algorithms. Some algorithms work quite well, such as HPI-Ncut algorithm, Min-Sum algorithm, and edge betweenness algorithm, while others are not. What causes such a difference? Fig.~\ref{fig_remove} may give us a clue. In this toy example, the original network is a two clusters SBM model with totally 2078 nodes and 3729 links. Fig.~\ref{fig_remove} shows the visualization of the top 10\% removed links of different algorithms. Please note that, the number of the red links in fig.~\ref{fig_remove}(b-f) are the same, namely, 373. However, comparing with edge betweenness and HPI-Ncut algorithm, much less of links between the two clusters are removed by EGP and CI algorithm, and more links are distributed among the left or the right cluster. Further more, comparing with edge betweenness algorithm, the links removed by HPI-Ncut algorithm mainly are distributed in the bridge part of the two clusters. This helps to partition the network into two disconnected clusters.

\begin{figure}
  \centering
  \includegraphics[width=16cm]{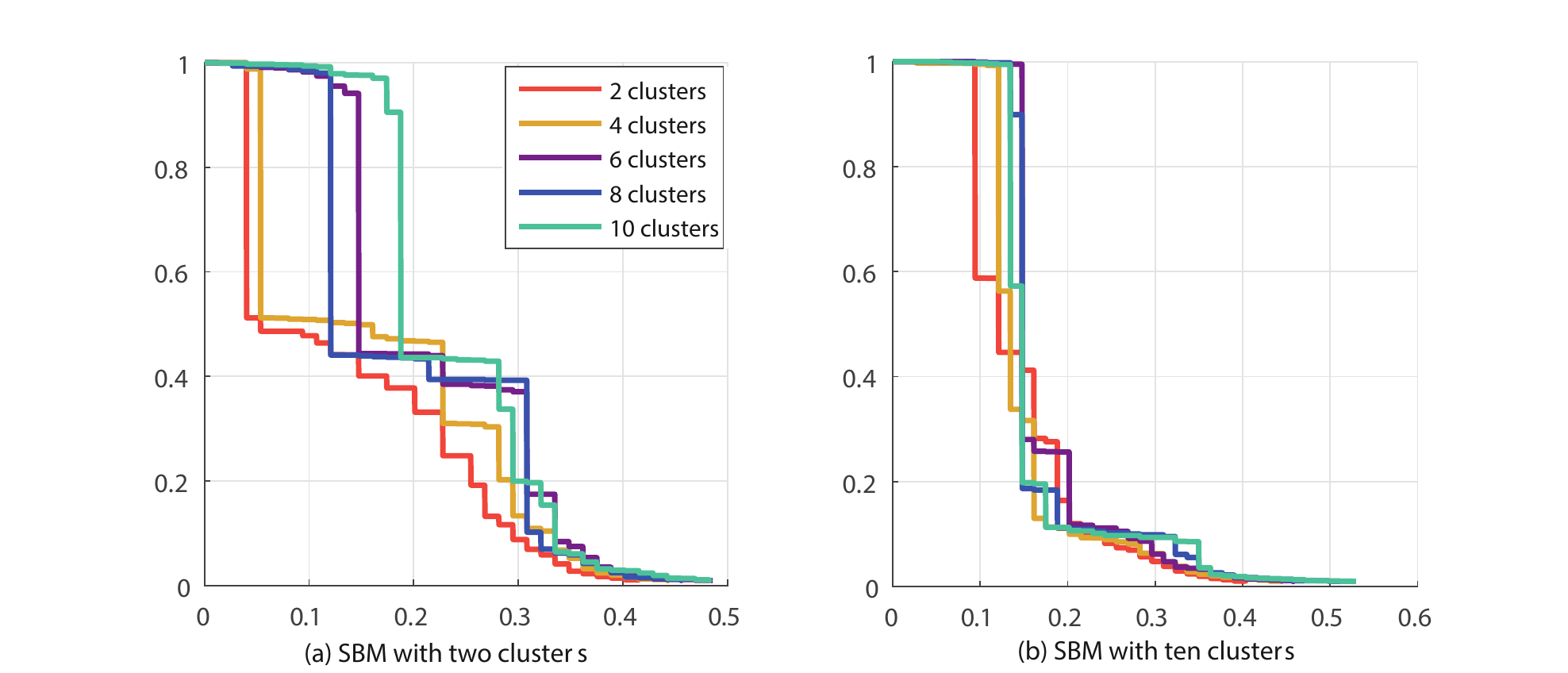}\\
  \caption{The size of the GCC of the networks versus link removing proportion, comparing of different quantities of target disconnected clusters in HPI-Ncut algorithm.}\label{fig_Ncut_k}
\end{figure}

In the previous sections, the default target number of the disconnected clusters in HPI-Ncut algorithm is set to 2. Fig.~\ref{fig_Ncut_k} shows the size of the GCC after targeted attack by HPI-Ncut with different target number of disconnected clusters, on the SBM network with two clusters and with ten clusters, respectively. Fig.~\ref{fig_Ncut_k} indicates that when the original networks contains less clusters, the target number of clusters in HPI-Ncut will greatly affect the size of GCC in the initial stage of the target attack, while, this influence will decline sharply in the later part of the attack process. However, the target number has a smaller impact on the attack performances of the HPI-Nuct when the original network contains much more clusters. Further more, when the target number of the disconnected clusters is set to 2, we can always obtain the optimal outcome on both networks. To conclude, we recommend to set the default target number of the disconnected clusters to 2 in HPI-Ncut algorithm.

\end{document}